\documentstyle[12pt,epsfig,lathuile]{article}
\begin{document}
\title{ 
SEARCHES FOR NEW PHYSICS AT HERA
}
\author{
Masahiro Kuze        \\
{\em Institute of Particle and Nuclear Studies, KEK, Tsukuba, 305-0801
  Japan} \\
(on behalf of the H1 and ZEUS Collaborations)
}
\maketitle
\baselineskip=14.5pt
\begin{abstract}
Recent results from searches for new physics at HERA are reviewed.
Exploiting the uniqueness of lepton-hadron collisions at high energy,
searches are performed for electron-quark resonant states (leptoquarks
or squarks in $R$-parity-violating supersymmetry) or excited states
of fermions.  New phenomena at a very high energy scale,
manifested at present energies as effective four-fermion contact interaction,
are also investigated, including cases with lepton-flavor violation.
Finally, the status of events with a high-energy lepton and missing
transverse momentum is presented, resulting in the most stringent constraint
on the flavor-changing neutral current $t$-$u$-$\gamma$ coupling which could
yield single-top production.

\end{abstract}
\baselineskip=17pt
\newpage
\section{Introduction}
HERA collides an electron or positron beam of 27.5~GeV and a
proton beam of 920~GeV (820~GeV until 1997), yielding 318~GeV (300~GeV)
center-of-mass energy ($\sqrt{s}$).  The square of the momentum transfer
($Q^2$) in deep inelastic scattering (DIS) can reach several times
$10^4~{\rm GeV}^2$, which means that the structure of the proton is probed
with a wavelength as small as one thousandth ($10^{-16}$~cm) of its radius.

During the years 1994-2000, each of the two collider experiments,
H1 and ZEUS, has collected approximately 110~$\rm pb^{-1}$ of $e^+p$
and 15~$\rm pb^{-1}$ of $e^-p$ data.  With these data, the experiments are
sensitive to rare processes with a cross section of a fraction of 
a picobarn, and searches for exotic signals of physics beyond the
Standard Model (SM) are extensively performed.
Possible signals of new physics include:
\begin{itemize}
\item Resonant states formed by electron-quark fusion.  Examples of
such states are leptoquarks or squarks (superpartners of quarks)
in supersymmetry (SUSY) with $R$-parity violation.  On-shell production
of such particles is possible for those with masses up to $\sqrt{s}$.
\item Physics at a much higher energy scale than the HERA energy could
cause a virtual effect in electron-proton scattering in the highest-$Q^2$
domain.  Such an effect is modeled as an effective four-fermion ($eeqq$) contact
interaction.  A variant of such a phenomenon is a lepton-flavor-violating (LFV)
interaction in which the incoming electron changes its flavor to
one of the higher-generation leptons ($\mu$ or $\tau$).
\item If the fermions (leptons and/or quarks) are composite rather than
elementary, excited states of fermions could be produced as a
consequence of the large momentum transfer, if the masses of such states
are below the HERA $\sqrt{s}$.  Such excited fermions could decay
back to the ground state (normal fermions) ``radiatively'' by emitting
a gauge boson ($\gamma$, $W$, $Z$ or gluon).
\end{itemize}
In the following sections, recent results are presented from both
collaborations on these topics.  Most of the results are preliminary.
\section{Leptoquarks}
Leptoquarks (LQs) are particles which carry both lepton (L) and baryon 
(B) numbers.
They appear in many unifying theories which try to establish fundamental
relations between leptons and quarks.
If LQs with a mass below the $\sqrt{s}$ of HERA exist, they could
be directly produced via a fusion between the electron and a quark
in the proton.
The production cross section depends on the unknown Yukawa coupling
$\lambda$ at the LQ-electron-quark vertex.
To evaluate the experimental results, the theoretical framework
of Buchm\"uller-R\"uckl-Wyler (BRW)\cite{BRW} is often used.
In this framework, LQs decay to an electron-quark pair with
a fixed branching ratio of either 1/2 or 1, depending on the LQ species.
The only other possible decay mode is into a neutrino-quark pair.

When the LQ decays to an electron-quark pair, the final state is identical
to SM neutral current (NC) DIS.  However, the angular distribution
is different.  NC DIS with $t$-channel photon exchange has a
$1/y^2$ angular distribution, where $y$ is the inelasticity variable
related to the electron decay angle in the $e$-$q$ rest frame, $\theta^*$,
with $\cos\theta^* = 1-2y$.
On the other hand, scalar LQs have flat $y$ distribution and vector
LQs have $(1-y)^2$ dependence.
Therefore, a cut in $y$ (or $\cos\theta^*$) is used to enhance the
LQ signal among the NC DIS background.

\begin{figure}[tb]
\epsfig{file=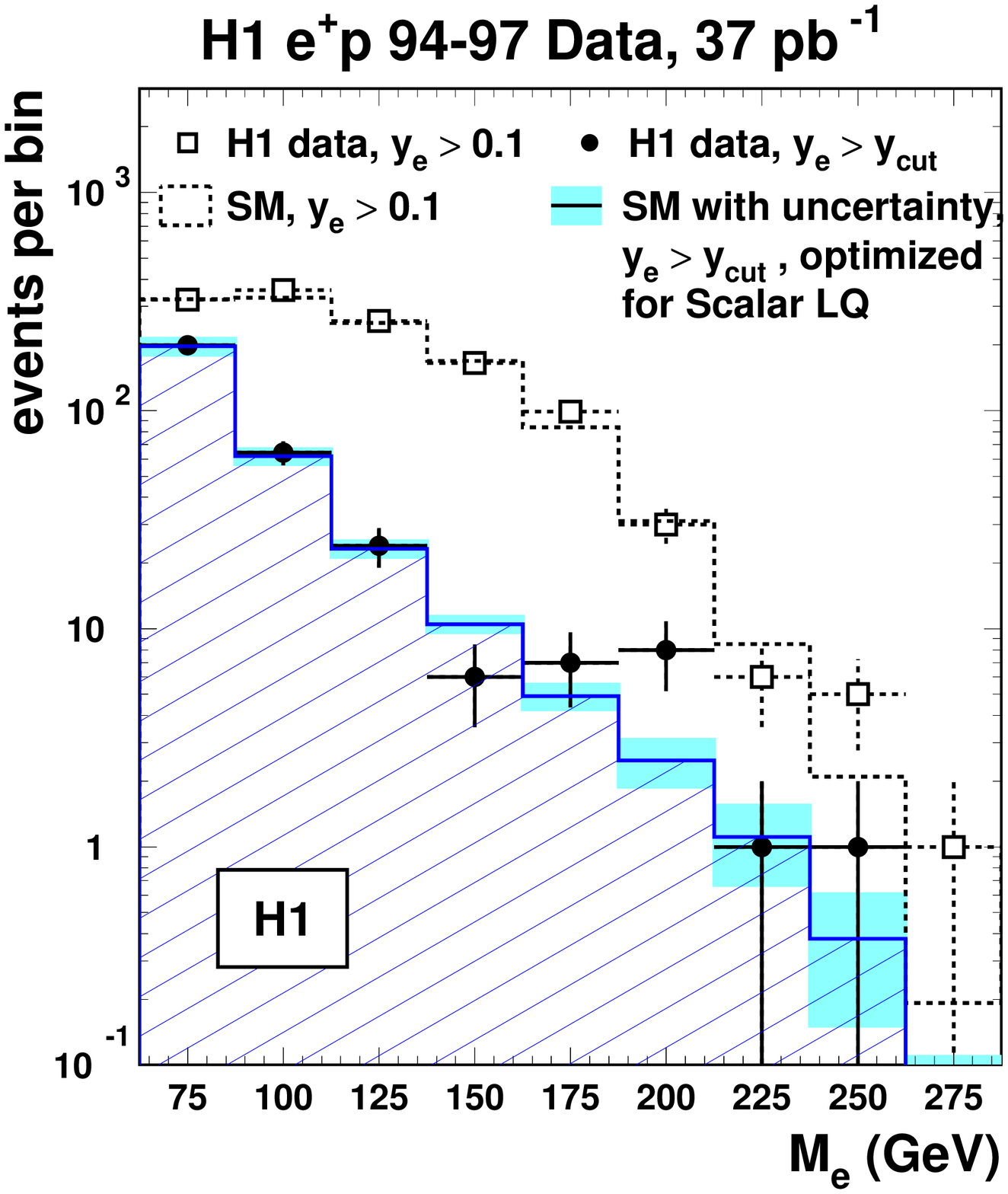,width=7.5cm}
\hspace{-0.5cm}
\epsfig{file=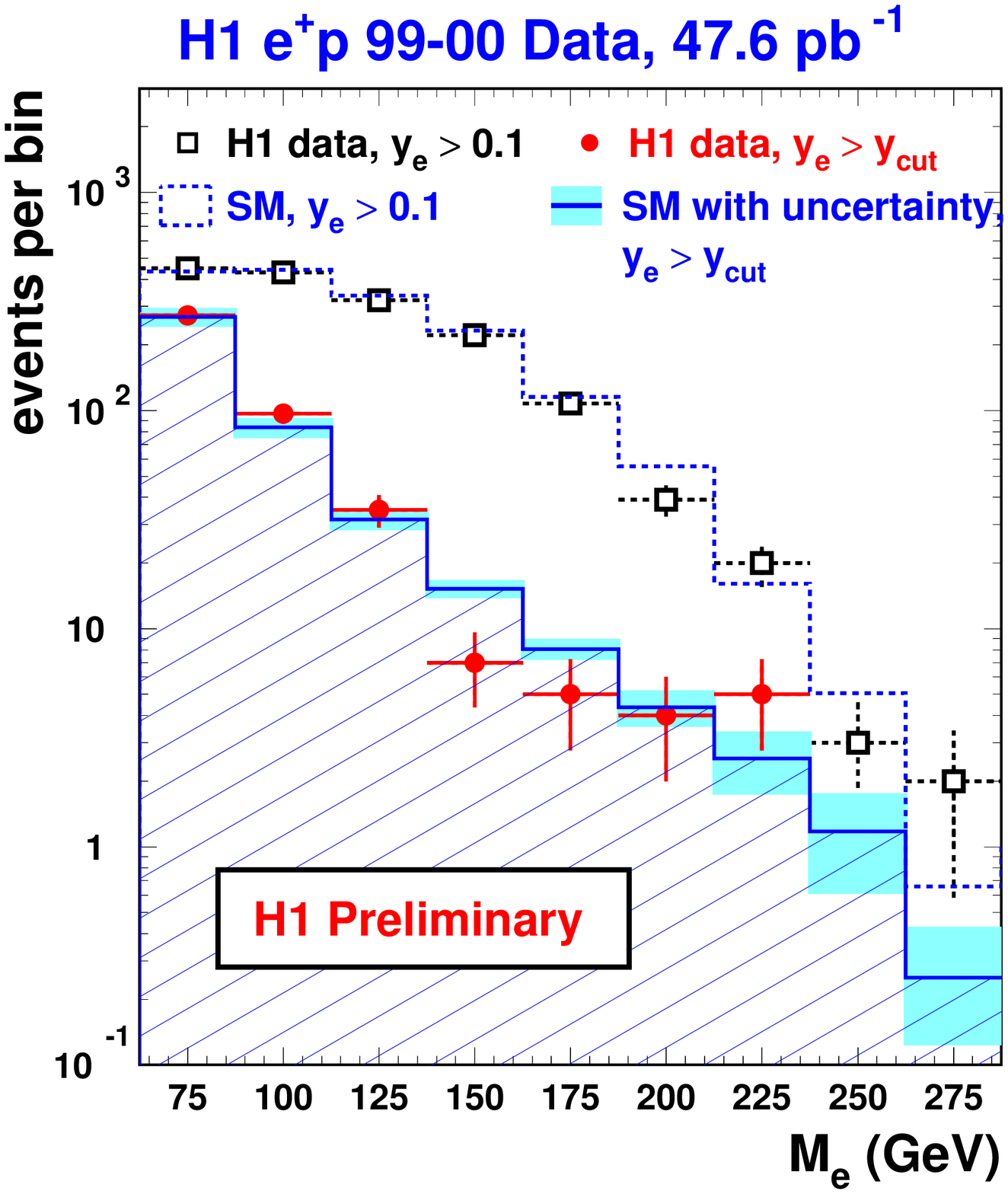,width=7.5cm}
 \caption{\it
Mass distribution of NC DIS candidates in H1 $e^+p$ data
from old (left) and new (right) running.  Distributions
before and after the $y$ cut designed to enhance a scalar LQ signal
are shown, and compared with SM predictions.
    \label{H1LQ} }
\end{figure}
Figure~\ref{H1LQ} shows the mass distribution of NC DIS candidates from 
the H1 experiment.  There was an excess of events after a $y$ cut in
the old $e^+p$ data\cite{PubH1LQ}, which is not confirmed by the new
$e^+p$ data taken at higher $\sqrt{s}$\cite{OsakaH1LQ}.
ZEUS also had a small excess in the high-$x$, high-$y$ region ($x$ is the
Bjorken scaling variable, related to the mass of the $eq$ system
via $M$=$\sqrt{sx}$) in the old data, which also is not confirmed by the
new data\cite{PubOsakaZEUSLQ}.
With the smaller amount of $e^-p$ data already analyzed, both
experiments have observed good agreement with SM predictions\cite{H1ZEUSFeq2}.

\begin{figure}[tb]
\begin{center}
\epsfig{file=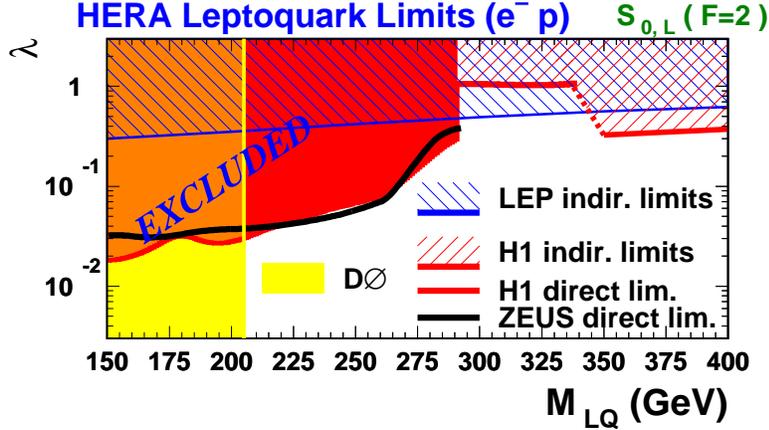,width=11cm}
 \caption{\it
Limits on the Yukawa coupling as a function of LQ mass for
$F$=2 LQ $S_0^L$.  Limits from the TeVatron and LEP are also shown.
    \label{Feq2lim} }
\end{center}
\end{figure}
With no evidence of a signal, limits are derived on the
Yukawa coupling as a function of the LQ mass, as shown in 
Fig.~\ref{Feq2lim}.
Note that $e^+p$ and $e^-p$ data have complementary sensitivity
to different LQ species; $e^+p$ collisions mainly probe the production
of $F$=0 LQs, which couple to a positron and a valence quark ($F$ is
the fermion number, defined as $F$=$L$+3$B$), and correspondingly
$e^-p$ for $F$=2.
In the figure, comparison is made with other collider results.
At the TeVatron, LQs can be pair-produced and thus the limit on mass
is independent of $\lambda$.
The limits from LEP come from hadronic cross section measurements
in which LQ exchange can produce virtual effects.

\begin{figure}[tb]
\begin{center}
\epsfig{file=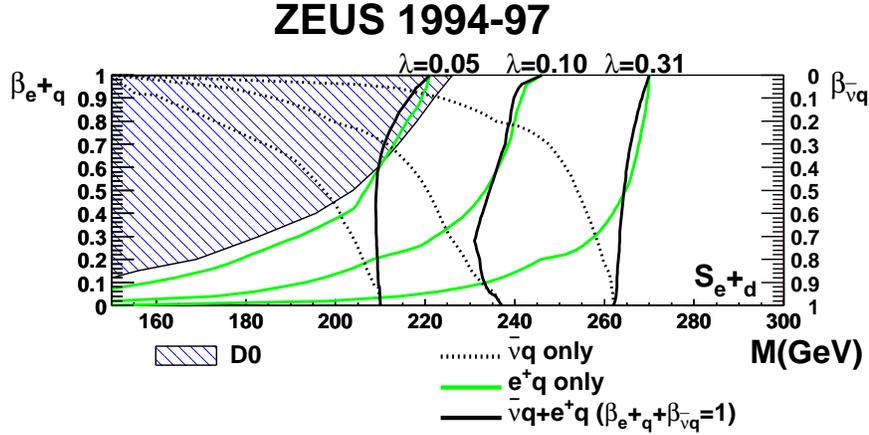,width=12cm}
 \caption{\it
Excluded region in the plane of decay branching ratio $\beta$
and LQ mass $M$, for a scalar LQ coupling to $e^+d$.  Limits
from an $e$-jet analysis only (left vertical axis), a $\nu$-jet
analysis only (right vertical axis) and also by combining the
two decay channels (assuming no other decay mode) are shown for
three values of the Yukawa coupling.  Also the limits from the TeVatron
are shown.  In all cases, the regions to the left of the curves are
excluded.
    \label{ZEUSnujet} }
\end{center}
\end{figure}
One can also consider a general case and treat the LQ branching ratio
to $eq$ as a free parameter, in contrast to the restrictions of the BRW
framework.  In Fig.~\ref{ZEUSnujet}, the search results in  both NC
(electron-jet resonance) and CC (neutrino-jet resonance) channels are
used, and combined limits are obtained assuming the
branching ratios to $eq$ and $\nu q$ decays sum up to one\cite{PubZEUSnujet}.
The limits in LQ mass for fixed values of $\lambda$ are largely
independent of how the branching ratios are shared, while the
TeVatron limits degrade largely when the LQ predominantly decays to $\nu q$.

\section{Squarks in supersymmetry with $R$-parity violation}
$R$-parity is a multiplicative quantum number defined as
$(-1)^{L+3B+2S}$, where $S$ is the spin of the particle.
It takes the value $+1$ for usual particles and $-1$ for
their superpartners.
In the Minimal Supersymmetric Standard Model (MSSM), $R$-parity is assumed
to be conserved.  The consequences are that superparticles are
always produced in pairs, and that the lightest
superparticle is stable.

However, the general supersymmetric and gauge-invariant superpotential
contains three terms which violate $R$-parity, two with $L$ violation
and one with $B$ violation\cite{RPV}.  Of particular interest at HERA
is the term $\lambda'_{ijk}L_i Q_j \bar D_k$, where $L$ and $Q$ are
left-handed lepton and quark doublet superfields and $\bar D$ is
a right-handed singlet superfield of down-type quarks. The indices
$i$, $j$ and $k$ denote their generation.  For each generation combination,
a new Yukawa coupling $\lambda'$ is introduced.

With a presence of non-zero $\lambda'_{1jk}$, production
of squarks is possible in $eq$ fusion at HERA, in exact analogy
to the scalar LQ production described before.
However, the decay mode of the squark has more varieties than
LQ decays in the BRW framework.  In addition to the $eq$ decay, 
a squark can decay to a quark and a gaugino with $R$-conserving
gauge couplings.  The gauginos can decay to a lepton ($e$ or $\nu_e$) and two
quarks with the same $\lambda'$ coupling, or to a lighter
gaugino and two fermions with gauge couplings which results
in further cascade decays.
As a consequence, there are many multi-jet and/or multi-lepton
final states with possible missing momentum,
which compete with the simple LQ-like decays.
It is worth noting that both $e^+$ and $e^-$ final states are
possible in the gaugino decay to $eqq$, which makes possible
a ``wrong sign'' electron search with respect to the charge of
the beam electron, a channel with very little background from
SM processes.

\begin{figure}[tb]
\begin{center}
\epsfig{file=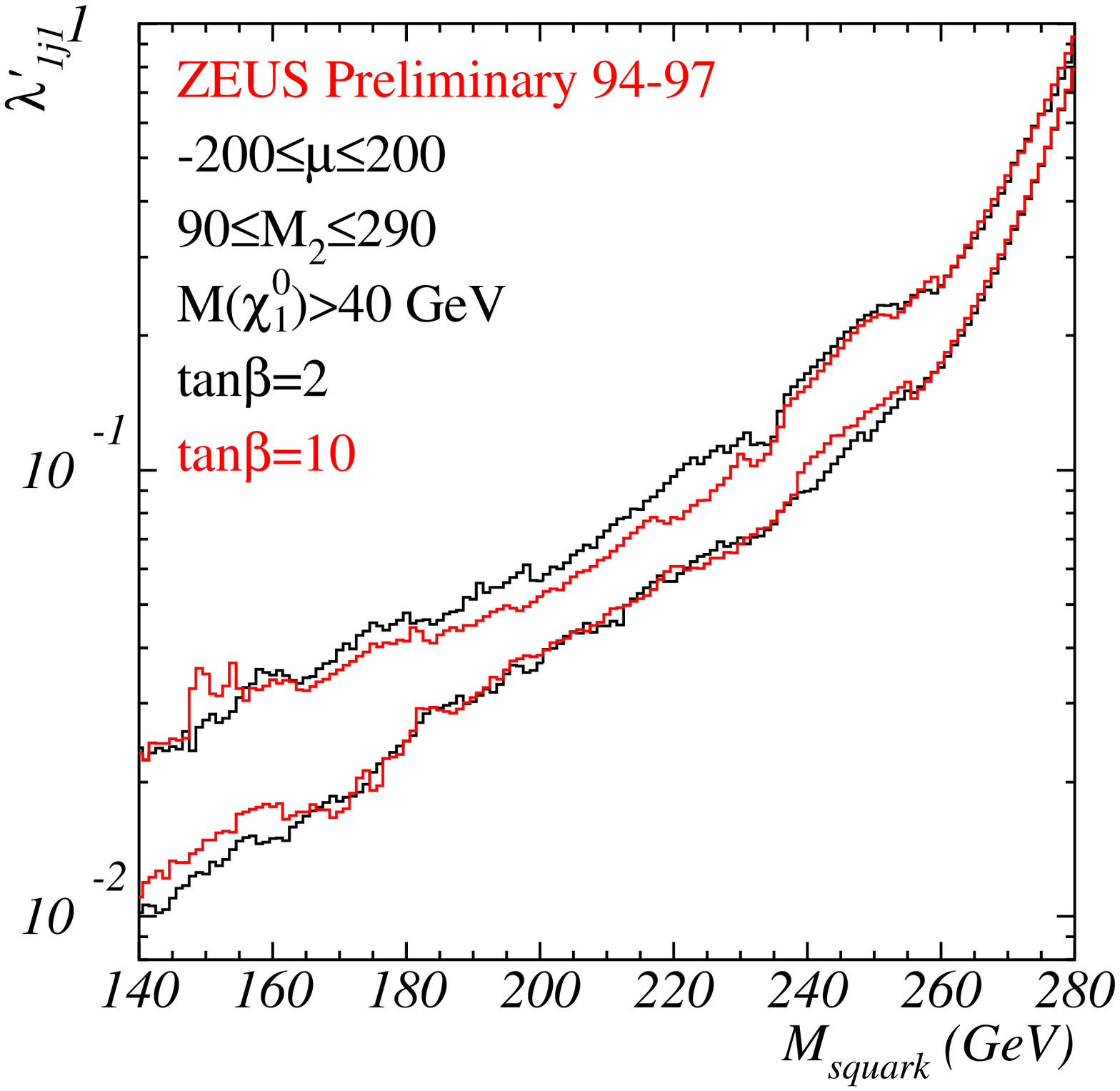,width=7.4cm}
\epsfig{file=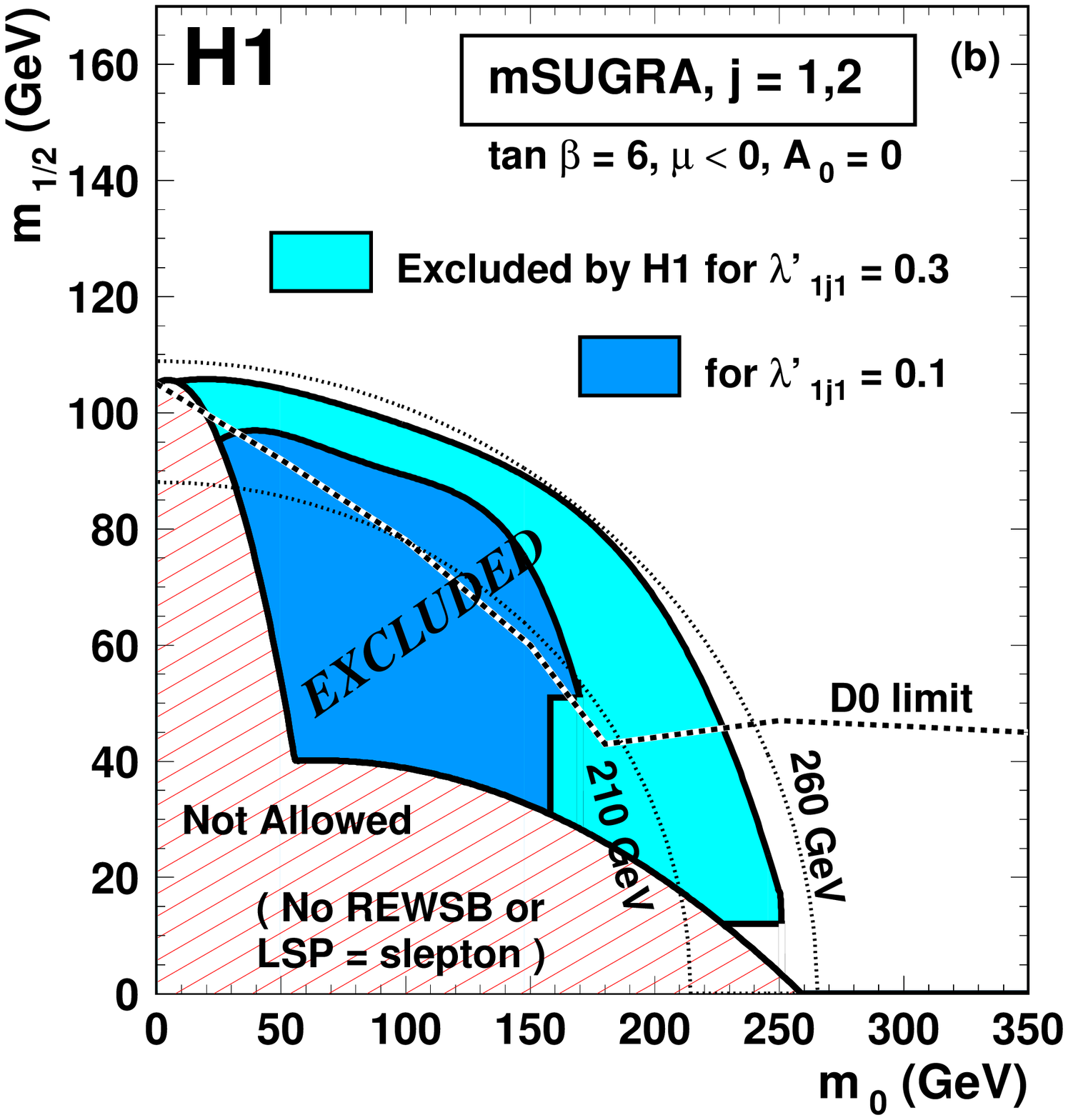,width=7.4cm}
 \caption{\it
On the left, ZEUS results for squarks in $R$-parity-violating SUSY are
presented as upper limits on the $\lambda'$ coupling as a function of squark mass.
For two values of $\tan \beta$, a scan in $\mu$ and $M_2$ is
performed, and the largest and smallest upper limits in the scan
parameter space are represented by the upper and lower curves.
On the right, H1 results in the mSUGRA scheme are presented, for the case of
$\tan \beta$=6.  For two values of the $\lambda'$ coupling, excluded region
in $(m_0, m_{1/2})$ plane is depicted, and compared with the TeVatron limit.
The contours for equal squark masses are also shown.
    \label{H1ZEUSSUSY} }
\end{center}
\end{figure}
Both H1 and ZEUS experiments have analyzed the $e^+p$ data taken
during 1994-97 (approx. 40~$\rm pb^{-1}$ per experiment) in
many decay channels, and no evidence of squark production was found.
The results can then be converted to constraints in SUSY parameter
space under various scenarios.
Figure~\ref{H1ZEUSSUSY} (left) shows the ZEUS results in the unconstrained MSSM
framework, in which squark masses are treated as parameters
independent of the MSSM parameters $\mu$, $M_2$ and $\tan \beta$.
For two fixed values of $\tan \beta$, a scan in the large parameter space of
$\mu$ and $M_2$ has been performed, and upper limits in $\lambda'_{1j1}$
have been obtained as a function of the squark mass\cite{ZEUSSUSY}.
The H1 collaboration has also derived limits in 
the mSUGRA framework, where the sparticle masses are determined from
other SUSY parameters and the number of free parameters become
minimal\cite{H1SUSY}.
The results are presented in Fig.~\ref{H1ZEUSSUSY} (right).
For fixed values of $\lambda'$ and $\tan \beta$, the excluded region in
SUSY parameter space ($m_0$, $m_{1/2}$) is shown.
For reasonably large values of the Yukawa coupling (even as small as 0.1), 
HERA has a sensitivity in a region not explored at the TeVatron,
although it should be mentioned that the TeVatron limits are independent
of $\lambda'$.
\section{Physics at very high energies}
\begin{figure}[tb]
\epsfig{file=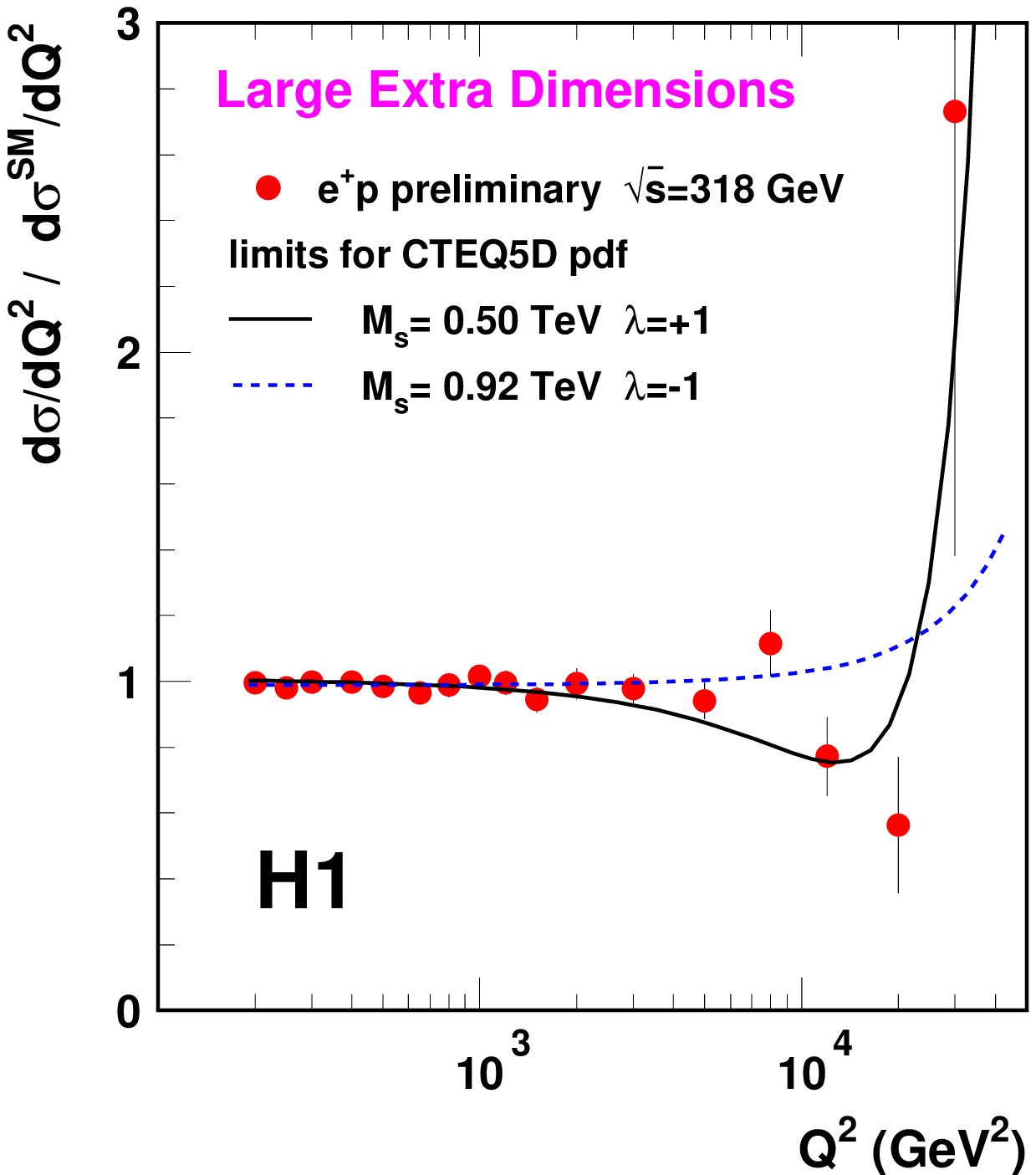,width=7.5cm}
\hspace{-0.5cm}
\epsfig{file=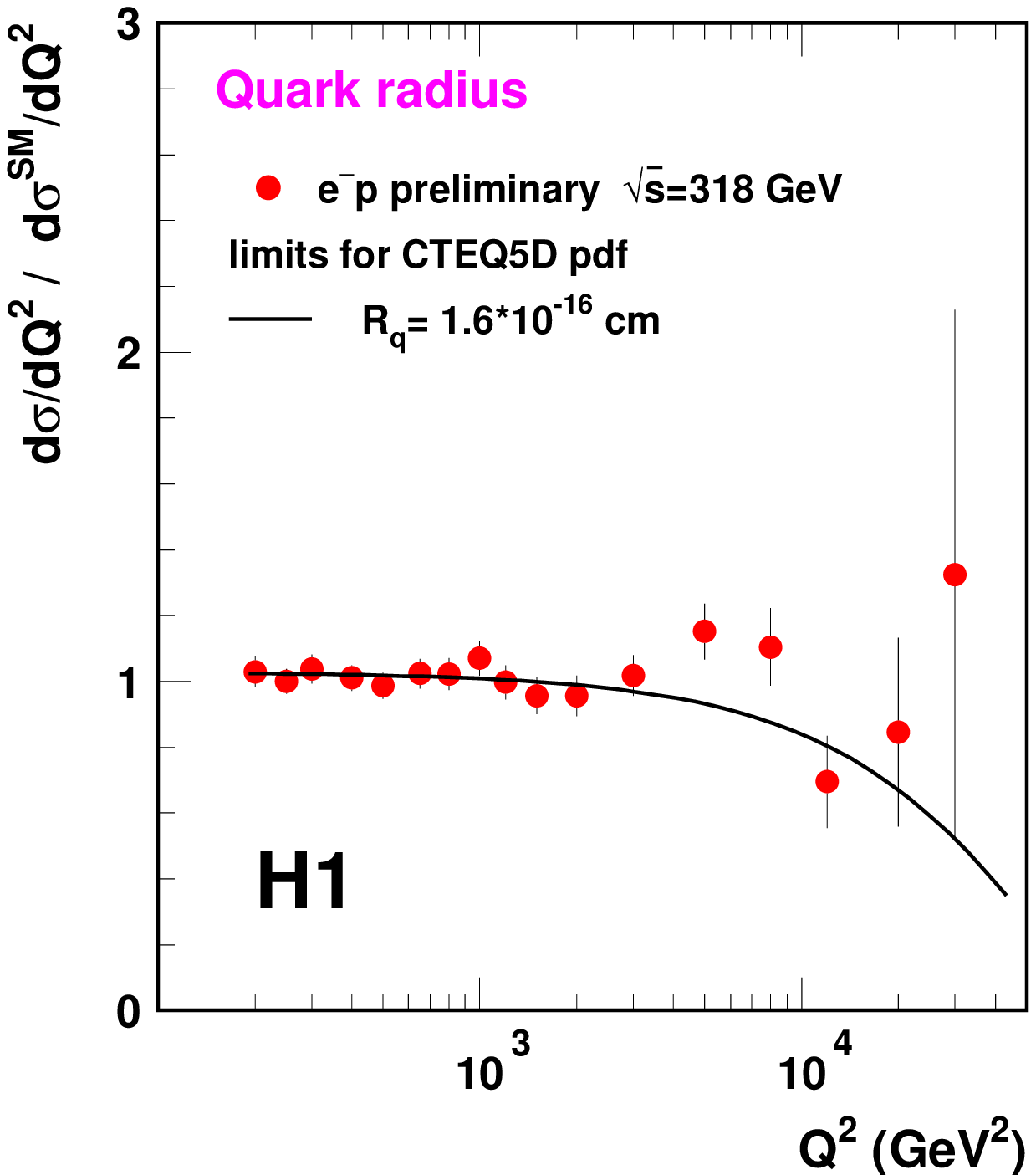,width=7.5cm}
 \caption{\it
Ratio of H1 $e^+p$ NC DIS cross section to the SM prediction
as a function of $Q^2$, overlaid with the prediction of
large extra dimensions corresponding to the mass scale
excluded at 95\% confidence level (CL) (left), and that of $e^-p$ overlaid with
the prediction of a quark radius corresponding to the value
excluded at 95\% CL (right).
    \label{H1CI} }
\end{figure}
New physics at an energy scale which is much larger than the $\sqrt{s}$ of
HERA could also be detected via its virtual effects at the highest-$Q^2$
region ($Q^2 \approx $ several $10^4$~GeV$^2$) which corresponds to
an $e$-$q$ interaction at a distance as short as 0.001~fm.
Such new physics processes are modeled in their low-energy limit as an
effective four-fermion contact interaction (CI), and the effects are
parameterized by a coupling constant, $g_{CI}$,
and an effective mass scale, $\Lambda$.

Both collaborations\cite{ZEUSCI,H1CI} tested the most common
$eeqq$ CI models using the NC DIS data at high $Q^2$.
Since the data are well described by the SM within statistical errors,
limits are derived for each model tested.  The limit on
$\Lambda$ depends on the chiral structure of the model, and ranges
up to 9~TeV when taking the usual coupling convention $g_{CI}^2$=$4\pi$.
The HERA limits are competitive and complementary to those from LEP and
the TeVatron, where the tests are made in $e^+e^- \to$hadrons and
Drell-Yan production, respectively.

Limits are also set on some specific models.  In models with large
(compactified) extra dimensions in which only the gravitational
force can propagate to the extra dimensions, the fundamental Planck mass scale
could be as low as the electroweak scale ($\approx$ TeV)\cite{ADD}.
In such models, high-energy particle collisions
get additional contributions due to the exchange of a Kaluza-Klein tower of
gravitons which modify the cross sections at the highest energies.
This causes CI-like effects at HERA in the highest-$Q^2$ region
with a main contribution from the interference of
graviton exchange and normal NC exchange.
These effects are characterized by an effective coupling
$\propto \lambda / M^4_s$, where $M_s$ is close to
the fundamental Planck scale and $\lambda = \pm 1$
determines whether the interference is constructive or destructive.
The results from H1\cite{H1CI} are shown in Fig.~\ref{H1CI} (left).
An effective mass scale $M_s$ of nearly 1~TeV is excluded at 95\% CL.

Another example is a test of classical form factor of quarks.
If one assumes that quark has a finite size and its charge is
distributed with a r.m.s. of $R$, the NC DIS cross
section gets a suppression factor $(1-R^2Q^2/6)^2$ which dumps the
cross section as $Q^2$ increases.  Radius of the order of $10^{-16}$cm
is excluded, as shown in Fig.~\ref{H1CI} (right)\cite{H1CI}.

\section{Lepton flavor violation}
\begin{figure}[tb]
      Table 1: {\it ZEUS limits (95\% CL upper limit)
on $\lambda_{eq_i}\lambda_{\tau q_j}/M_{LQ}^2$
$(10^{-4}~GeV^{2})$ for $F$=0 LFV leptoquarks mediating the $eq_i
\leftrightarrow \tau q_j$ transition (bold numbers in the bottom of each
cell).  Each row corresponds to a $(q_i, q_j)$
generation combination and each column corresponds to a LQ species.
The numbers in the middle of each cell are the best limit from other
experiments.  The cases where the ZEUS limit is the
most stringent or comparable to the best limit within a factor of 2
are enclosed in a box.  The * shows the cases where only the top
quark can participate.
Similar tables exist for $F$=2 LQs, and for the $e$-$\mu$
transition.}
\begin{center}
\vskip 0.1 in
\epsfig{file=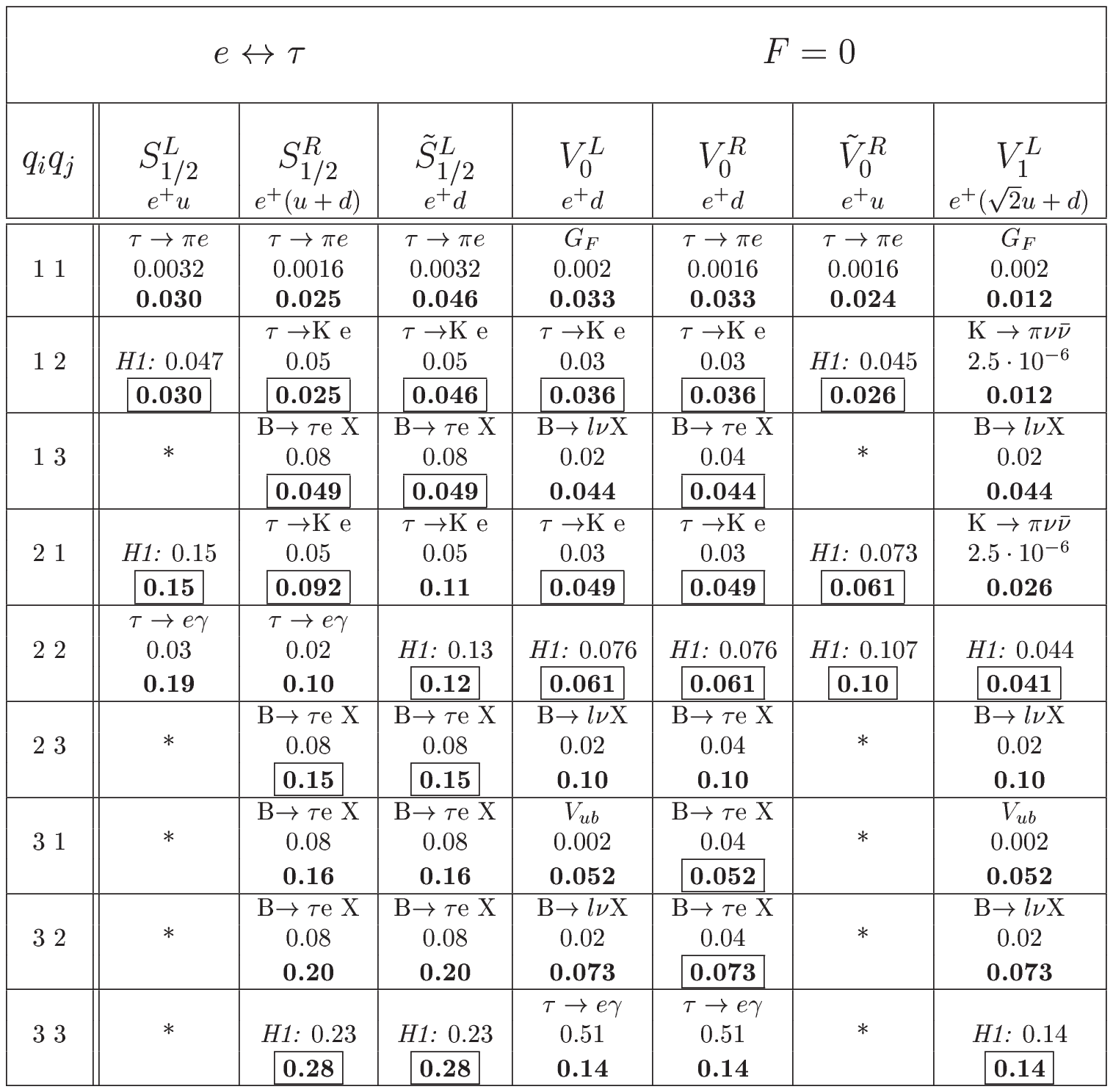,width=10.6cm}
\end{center}
\end{figure}

\begin{figure}[tb]
\epsfig{file=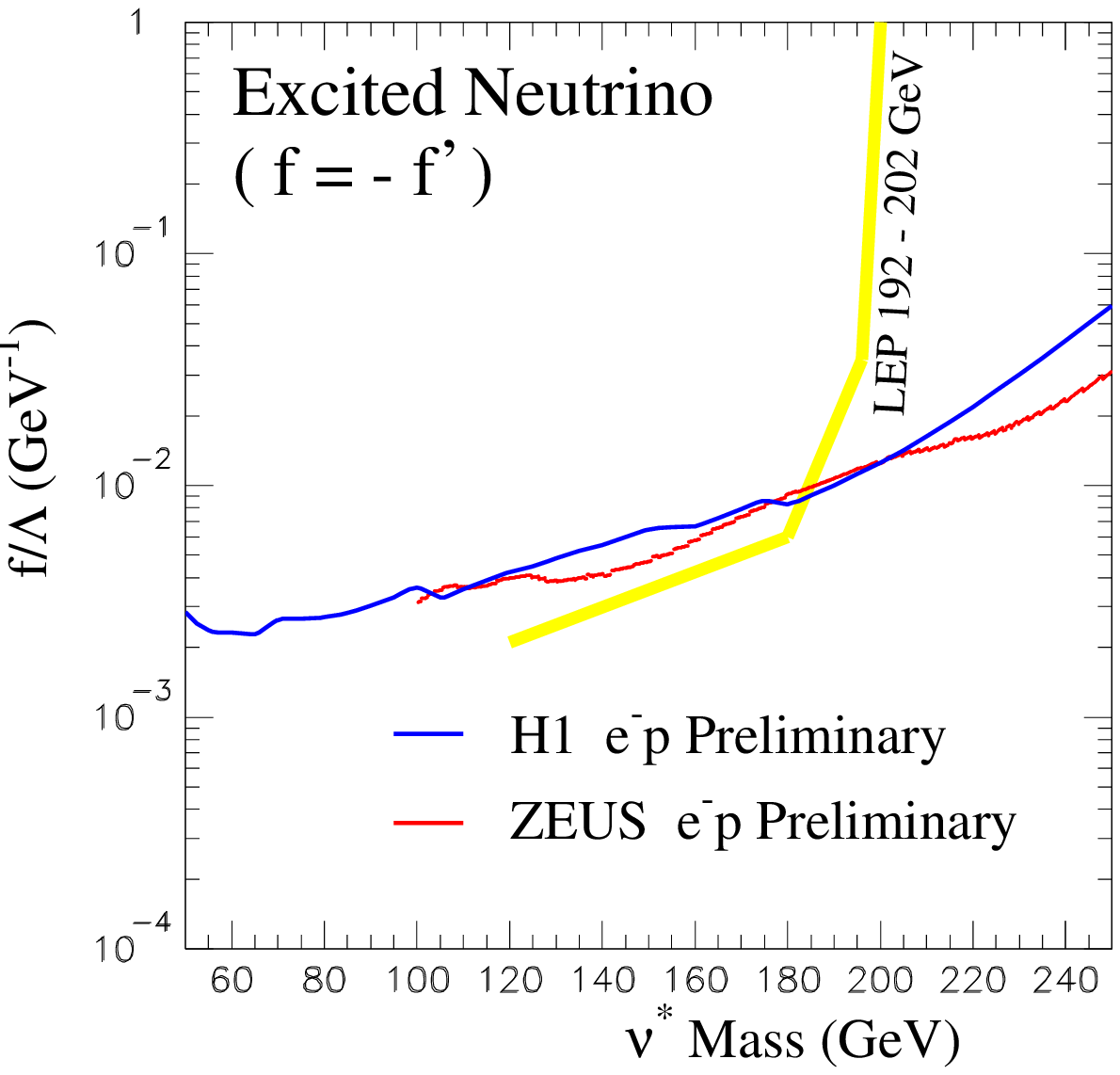,width=7.9cm}
\hspace{-1cm}
\epsfig{file=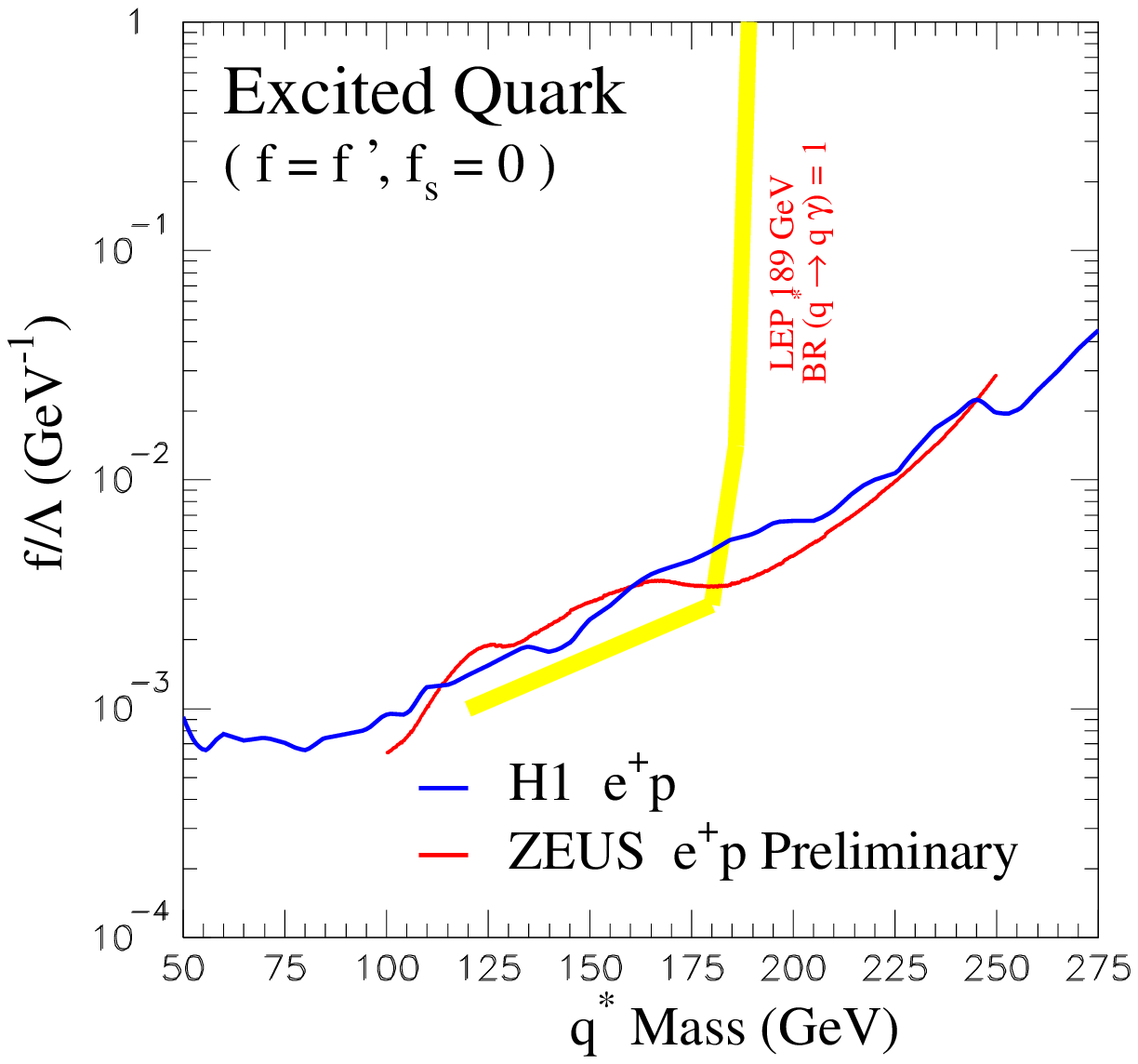,width=7.9cm}
 \caption{\it
Upper limits on $f/\Lambda$ as a function of the excited-fermion mass
for $\nu^*$ with an assumption $f$=$-f$' (left) and for $q^*$
with an assumption $f$=$f'$, $f_s$=0 (right), compared to the
limits from LEP.
    \label{EXF} }
\end{figure}

In the SM, lepton flavor is conserved in every interaction.
However, it is not regarded as a fundamental symmetry, and new
physics could naturally entail the inter-generation transition of leptons.
HERA is well suited for searching for electron-muon or electron-tau
LFV processes by colliding an electron with a proton at a very short distance.
The final state will be an $ep \to \mu (\tau) X$ event which has a
striking signature of a high-momentum muon or tau balanced by a hadronic
system, which has very little background from SM processes.

Both experiments searched for such events in the 1994-97 data, with no
candidates\cite{PubH1LQ, ZLFV}.  An example of the limits are expressed in
Table~1 in the context of heavy ($ \gg \sqrt{s}$) LFV leptoquarks mediating
the $eq \leftrightarrow l q'$
transition, for each LQ species and $(q,q')$ generation combination.
It can be seen that limits from low-energy experiments, such as rare decays,
are very stringent for transitions involving first-generation quarks, but
HERA has higher (or unique) sensitivity when the LFV transition involves
higher-generation quarks.
\section{Excited fermions}
\begin{figure}[tb]
\hspace{1.5cm}
\epsfig{file=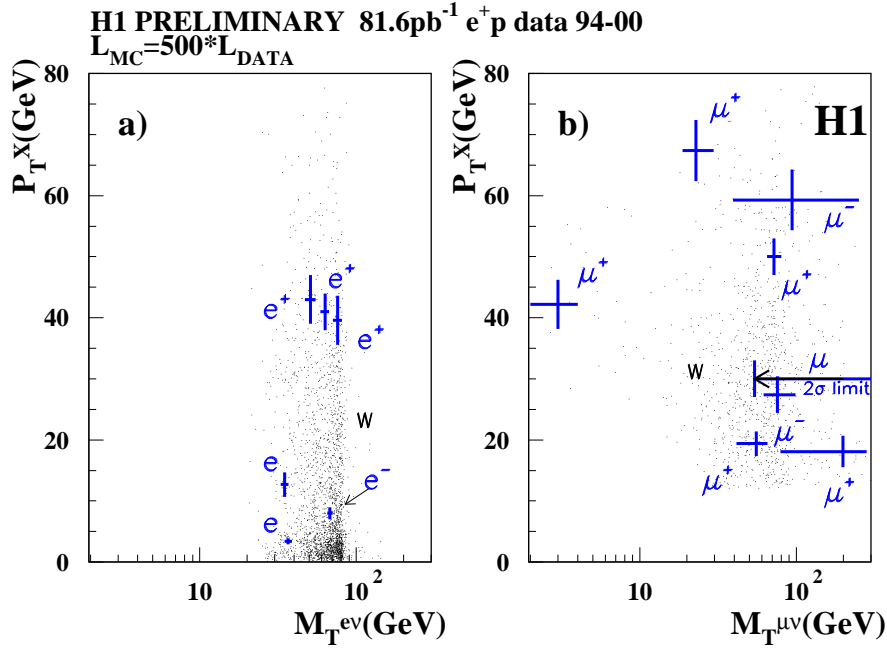,width=13cm}
 \caption{\it
Distribution of H1 events with isolated lepton and missing $P_T$
in the plane of hadronic $P_T$ and transverse mass.  The left plot
shows the electron channel and the right plot is for the muon channel.
A Monte Carlo distribution of $W$ production, with 500 times the luminosity
of the data, is also overlaid.
    \label{H1hpt} }
\end{figure}
If fermions are not elementary but have substructure,
their excited states could be produced in a high-energy collision.
At HERA, excited electrons and quarks could be produced mainly via
$t$-channel photon exchange, and excited neutrinos via $W$ exchange.
Once produced, they can decay to a normal fermion by radiating off
gauge bosons.  Decays into $\gamma$, $W$ or $Z$ are considered here.

To measure the experimental sensitivity, the theoretical framework
of Hagiwara, Komamiya and Zeppenfeld\cite{HKZ} is conventionally used,
in which the production cross section depends on the relative coupling
strengths $f$, $f'$ and $f_s$ for the SU(2), U(1) and SU(3) gauge groups,
respectively, and a common compositeness scale $\Lambda$.
By assuming relations between the couplings, the decay branching ratios
to different gauge bosons are fixed and the cross section depends
only on the parameter $f/\Lambda$.

Both collaborations searched for excited fermions in various decay
modes (including hadronic and leptonic decays of $W$, $Z$) in the
1994-97 $e^+p$ data and 1998-99 $e^-p$ data, with no evidence
of a signal\cite{EXF}.  It should be noted that $e^-p$ collisions provide
a much larger cross section for excited-neutrino production than $e^+p$ because
of the larger $u$-quark density in the proton and the helicity suppression
in $e^+p$ due to $W$-exchange.

Figure~\ref{EXF} shows the upper limits obtained on $f/\Lambda$ as
a function of the excited-fermion mass, for $\nu^*$ and $q^*$.
HERA experiments have a unique sensitivity beyond the LEP2 center-of-mass
energy.  Stringent limits on $q^*$ are obtained from the TeVatron,
but the production at the TeVatron takes place via the strong coupling $f_s$.
HERA limits, on the other hand, are sensitive to the electroweak couplings
$f$ and $f'$ and are thus complementary.

\section{$W$ and single-top production}
\begin{table}
\setcounter{table}{1}
\centering
\caption{ \it Summary of isolated high-$P_T$ leptons from H1 (top)
and ZEUS (bottom). For two values of cuts in $P_T^X$, the observed and
expected numbers of events are shown.  The numbers in parenthesis are the
W contribution to the total SM expectation.  For H1, the error in the expected
number includes the systematic error.
}
\vskip 0.1 in
\begin{tabular}{|c|c|c|} \hline
H1 preliminary & Electrons & Muons \\
1994-2000 $e^+p$ 82~pb$^{-1}$ & Observed/expected (W) & Observed/expected (W) \\
\hline
 $P_T^X > 25$~GeV   & $ 3~/~1.05 \pm 0.27~(0.83)$ & $ 6~/~1.21 \pm 0.32~(1.01) $ \\
\hline
 $P_T^X > 40$~GeV   & $ 2~/~0.33 \pm 0.10~(0.31)$ & $ 4~/~0.46 \pm 0.13 (0.43) $ \\
\hline
\end{tabular}
\vskip 0.1 in
\begin{tabular}{|c|c|c|} \hline
ZEUS preliminary & Electrons & Muons \\
1994-2000 $e^\pm p$ 130~pb$^{-1}$ & Observed/expected (W) & Observed/expected (W) \\
\hline
 $P_T^X > 25$~GeV   & $ 1~/~1.14 \pm 0.06~(1.10)$ & $ 1~/~1.29 \pm 0.16~(0.95) $ \\
\hline
 $P_T^X > 40$~GeV   & $ 0~/~0.46 \pm 0.03~(0.46)$ & $ 0~/~0.50 \pm 0.08~(0.41) $ \\
\hline
\end{tabular}
\label{pttab}
\end{table}
\begin{figure}[tb]
\hspace{1cm}
\epsfig{file=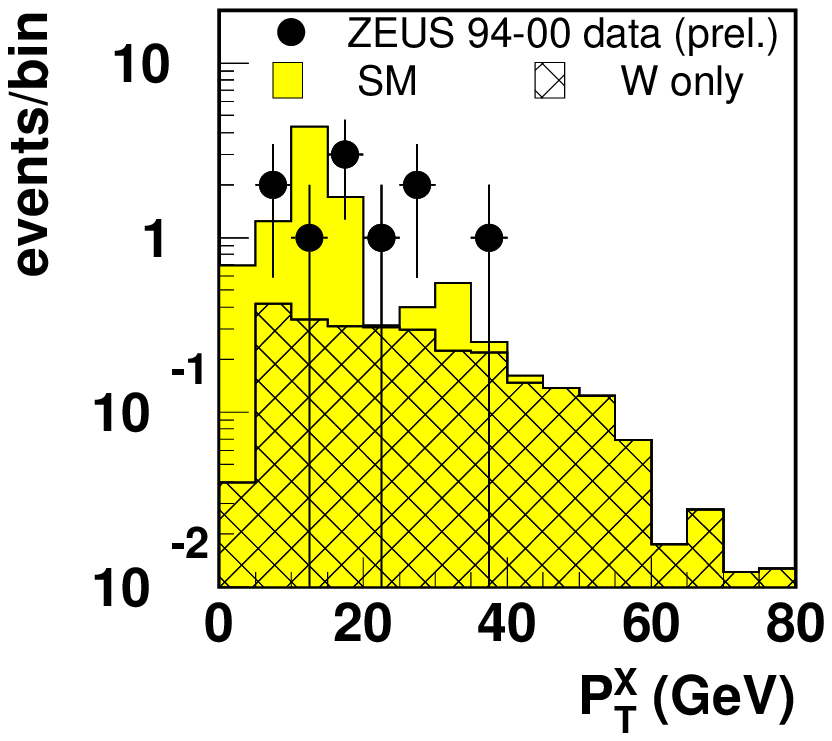,width=6.5cm}
\epsfig{file=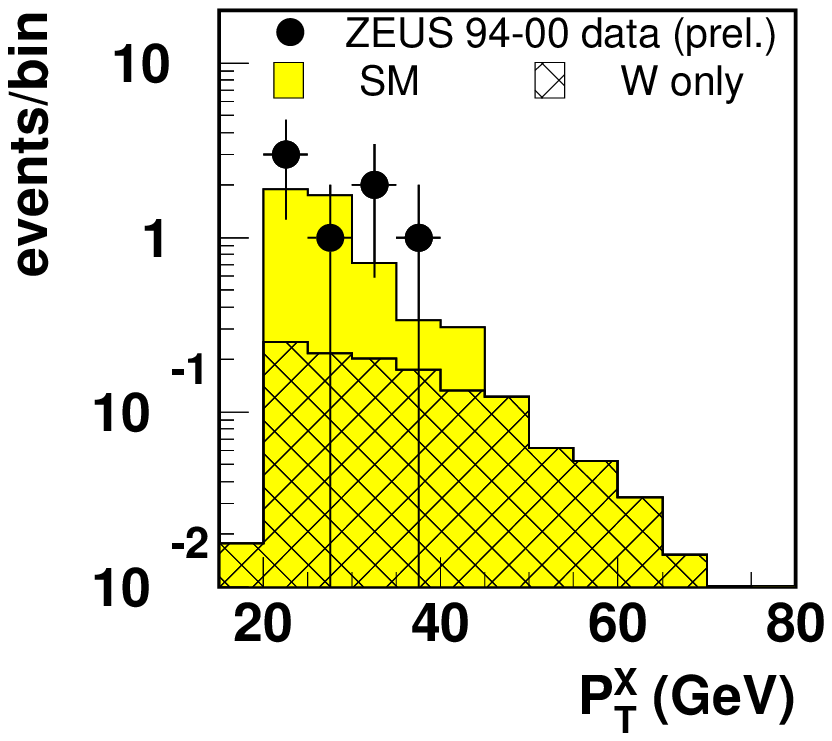,width=6.5cm}
 \caption{\it
      $P_T^X$ distribution of
ZEUS events in the electron (left) and muon (right) channels.
The shaded region in the MC expectation is the $W$ component.
    \label{ZEUSPTX} }
\end{figure}
\begin{figure}
\vspace{1cm}
\hspace{1cm}
\epsfig{file=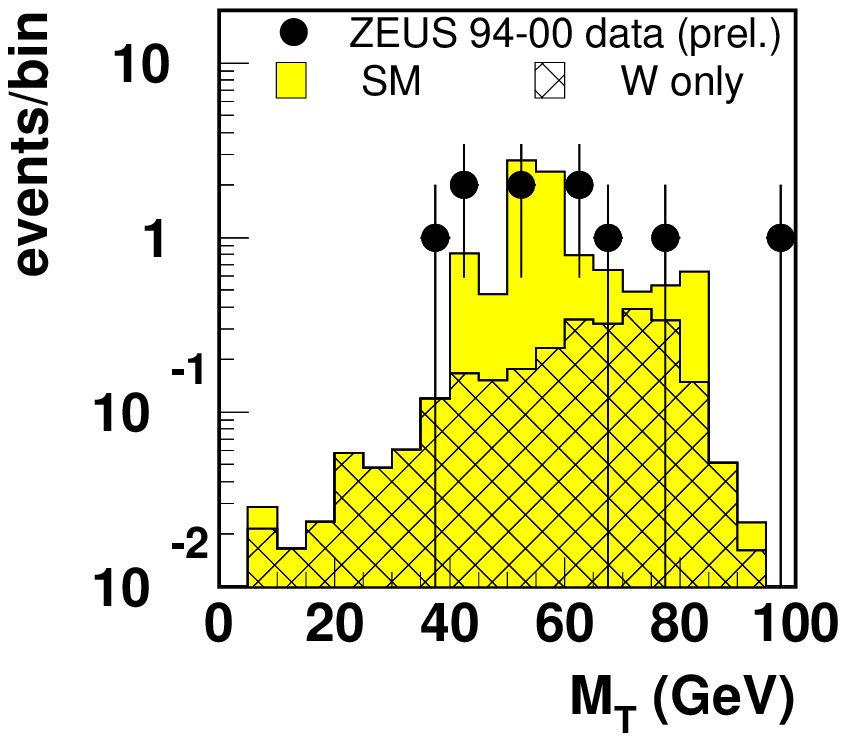,width=6.5cm}
\epsfig{file=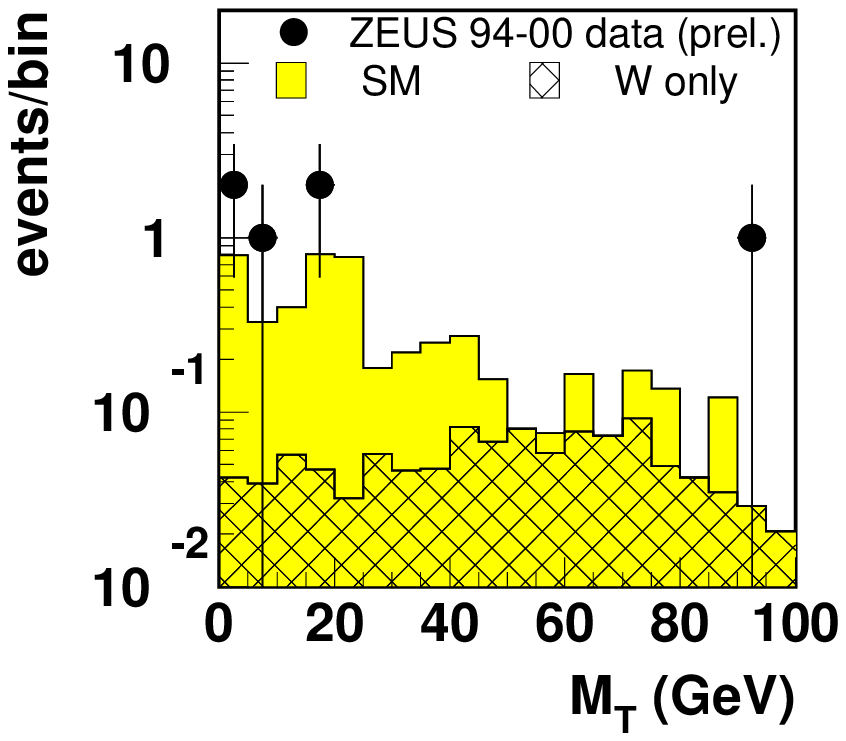,width=6.5cm}
 \caption{\it
      Transverse-mass distribution of ZEUS events in
the electron (left) and muon (right) channels.
The shaded region in the MC expectation is the $W$ component.
    \label{ZEUSMT} }
\end{figure}
Events with a high transverse momentum (high-$P_T$) lepton and missing $P_T$
at HERA have drawn attention in recent years due to the excess of such events
reported by the H1 experiment.  When the region with large $P_T^X$ (the transverse
momentum of the remaining hadronic system) is selected, the prediction
from the SM is dominated by single-$W$ production, which has a total
cross section of about 1~pb.

Figure~\ref{H1hpt} shows the results from H1 presented in summer 2000,
from the $e^+p$ data collected up to middle of 2000\cite{H1osakaW}.
Events are plotted in $P_T^X$ and $M_T$, the transverse mass between the
lepton and missing $P_T$.
There is no acceptance for muon events at low $P_T^X$ due to the
requirement of missing $P_T$ measured in the calorimeter.
The observed and expected numbers of events are shown in Table~\ref{pttab}
(top).  For $P_T^X > 25$~GeV, nine electron or muon events were observed
while $2.3 \pm 0.6$ are expected from the SM, dominated by $W$ production.

ZEUS had observed good agreement with SM predictions for these
event topologies up to the results presented in summer
2000\cite{ZEUSosakaW}, which
included data from 1994-99.  In this conference, the results
were updated for the first time using all of the data collected in 2000.

\begin{figure}[tb]
\epsfig{file=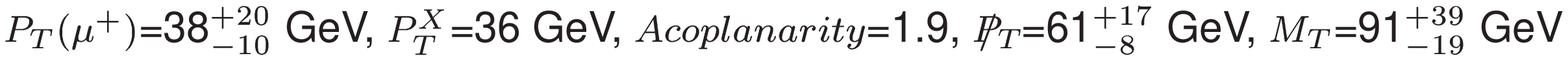,width=14.5cm}
\vskip 0.3cm
\epsfig{file=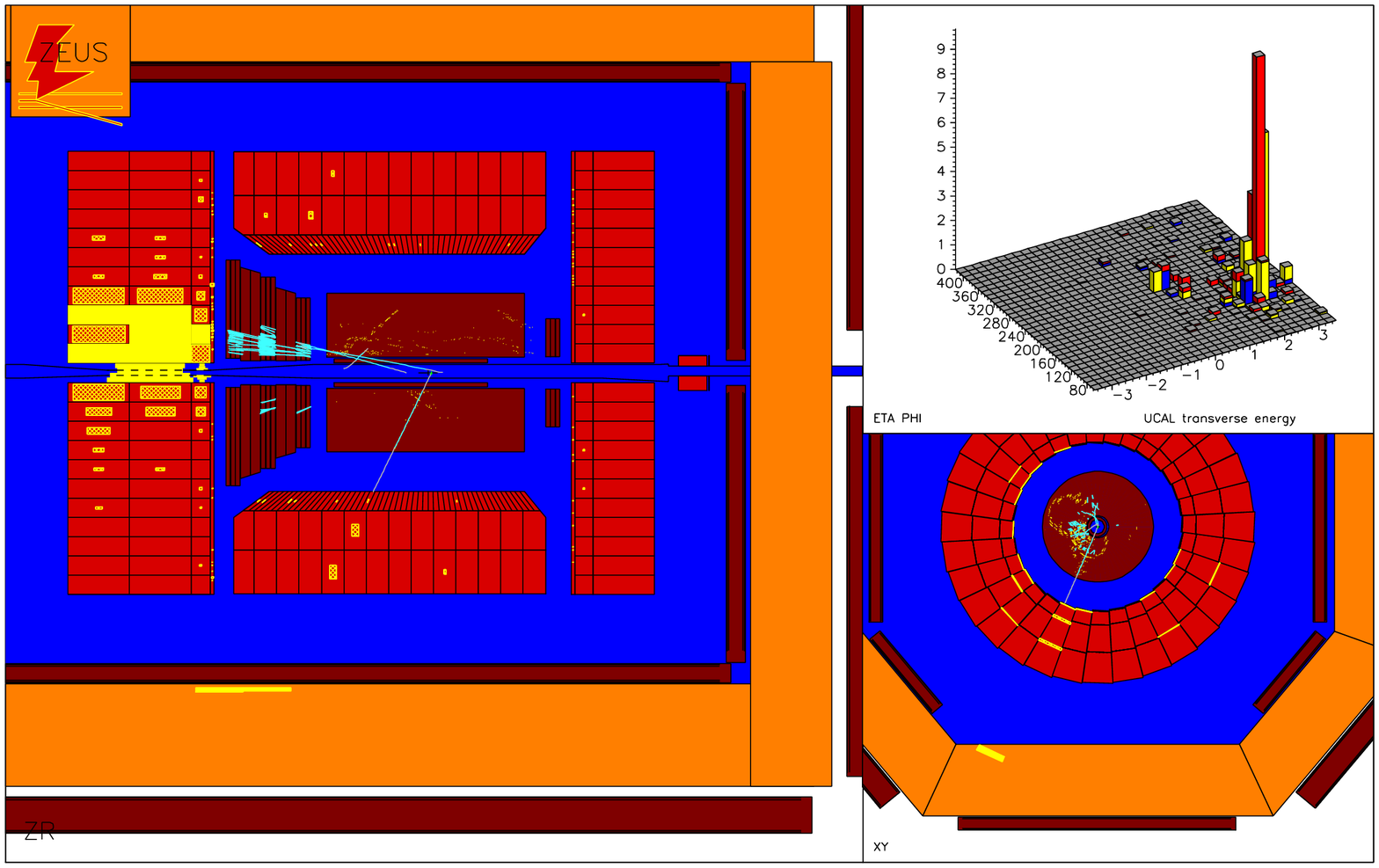,width=14.5cm}
 \caption{\it
      The ZEUS muon event from 2000 which passed the $P_T^X > 25~GeV$ cut.
    \label{nicemuon} }
\end{figure}
Figures~\ref{ZEUSPTX} and \ref{ZEUSMT} show the distributions of
$P_T^X$ and transverse mass for events with calorimeter
$P_T > 20$~GeV and an isolated track with $P_T^{track} > 10$~GeV.
Ten electron events and seven muon events were selected, with the
SM expectation being $11.0 \pm 1.6$ and $5.4 \pm 0.7$, respectively.
Both the event rate and the distribution of the kinematic variables
are in good agreement with SM expectations, which are dominated by
NC DIS for electrons and $\gamma \gamma \to \mu \mu$ for muons.

By further applying cuts to enhance the $W$ component in the
expectation, similar to the ones used in the H1 analysis, the numbers of events
are compared with the SM predictions in Table~\ref{pttab} (bottom) for the large-$P_T^X$ region.
In the ZEUS case, there is no excess of events observed.
There is somewhat larger acceptance for $W$ in the H1 analysis
due to the wider polar-angle coverage, but it is worth mentioning
that all of the observed H1 events are in the polar-angle range where
ZEUS has acceptance.
The muon event surviving the $P_T^X > 25$~GeV cut is from the 2000
data, and its event display is shown in Fig.~\ref{nicemuon}.

An event with a lepton, missing $P_T$ and a large $P_T^X$ is a 
typical signature for a top quark decaying via $t \to bW \to bl\nu$.
Therefore, the results of the searches described above can be applied
to single-top production, $ep \to etX$.  It is a flavor-changing
neutral current (FCNC) process and highly suppressed in the SM.
Any observation of single top production at HERA would be a signal
for physics beyond the SM.

\begin{figure}[tb]
\hspace{2cm}
\epsfig{file=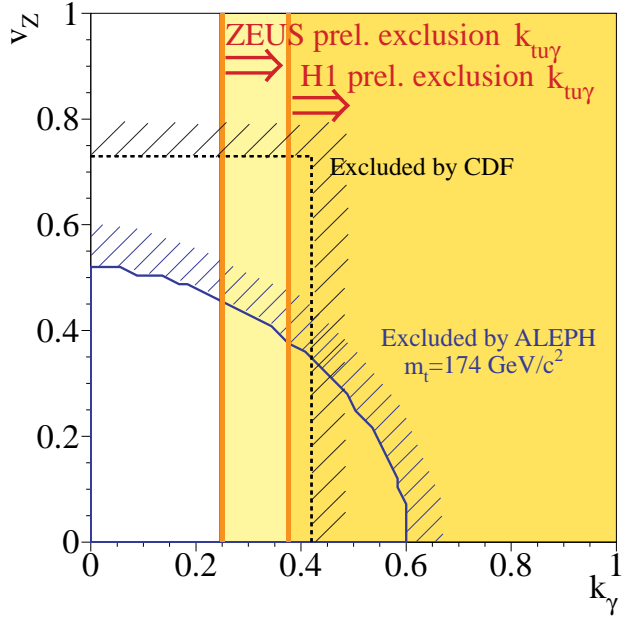,width=10cm}
 \caption{\it
      Limits at 95\% CL on FCNC magnetic coupling at the photon vertex,
$k_\gamma$, and vector coupling at the $Z$ vertex, $v_Z$, from LEP, TeVatron
and HERA.  The HERA limits apply only to $k_{tu\gamma}$.
    \label{FCNC} }
\end{figure}

The resulting limits\footnote{In the H1 single-top search, the hadronic
decay of $W$ is also used.}, together with those from LEP\cite{ALEPHW} and the
TeVatron, are expressed in Fig.~\ref{FCNC} in the plane of
two couplings, $k_\gamma$, the magnetic coupling at the photon-top-quark
vertex and $v_Z$, the vector coupling at the $Z$-top-quark vertex.
Here the quark can be generally $u$ or $c$.
At LEP, the limits come from a single-top search $e^+e^- \to tq$ and
the TeVatron limits are from a rare top decay search $t \to q\gamma, qZ$.
Both processes can constrain the $\gamma$ and $Z$ couplings, and the limits
apply to both $u$ and $c$ quarks since they are in the final state.
At HERA, the contribution from $Z$-exchange in the $t$-channel
is suppressed due to the large mass in the propagator, so the cross
section is dominated by the photon coupling $k_\gamma$.
Moreover, since the $u$ quark dominates the parton distribution in the proton
at large $x$, HERA is most sensitive to the $tu\gamma$ coupling.
It can be seen that the HERA limits on $k_{tu\gamma}$ are more stringent
than those from LEP and the TeVatron.
Note that the cross section dependence on the top mass uncertainty
(5~GeV) is about 20\% at HERA\cite{Belyaev}, while it is larger at
LEP where the available energy is close to the threshold.
\section{Summary and prospects}
\begin{figure}[tb]
\hspace{2cm}
\epsfig{file=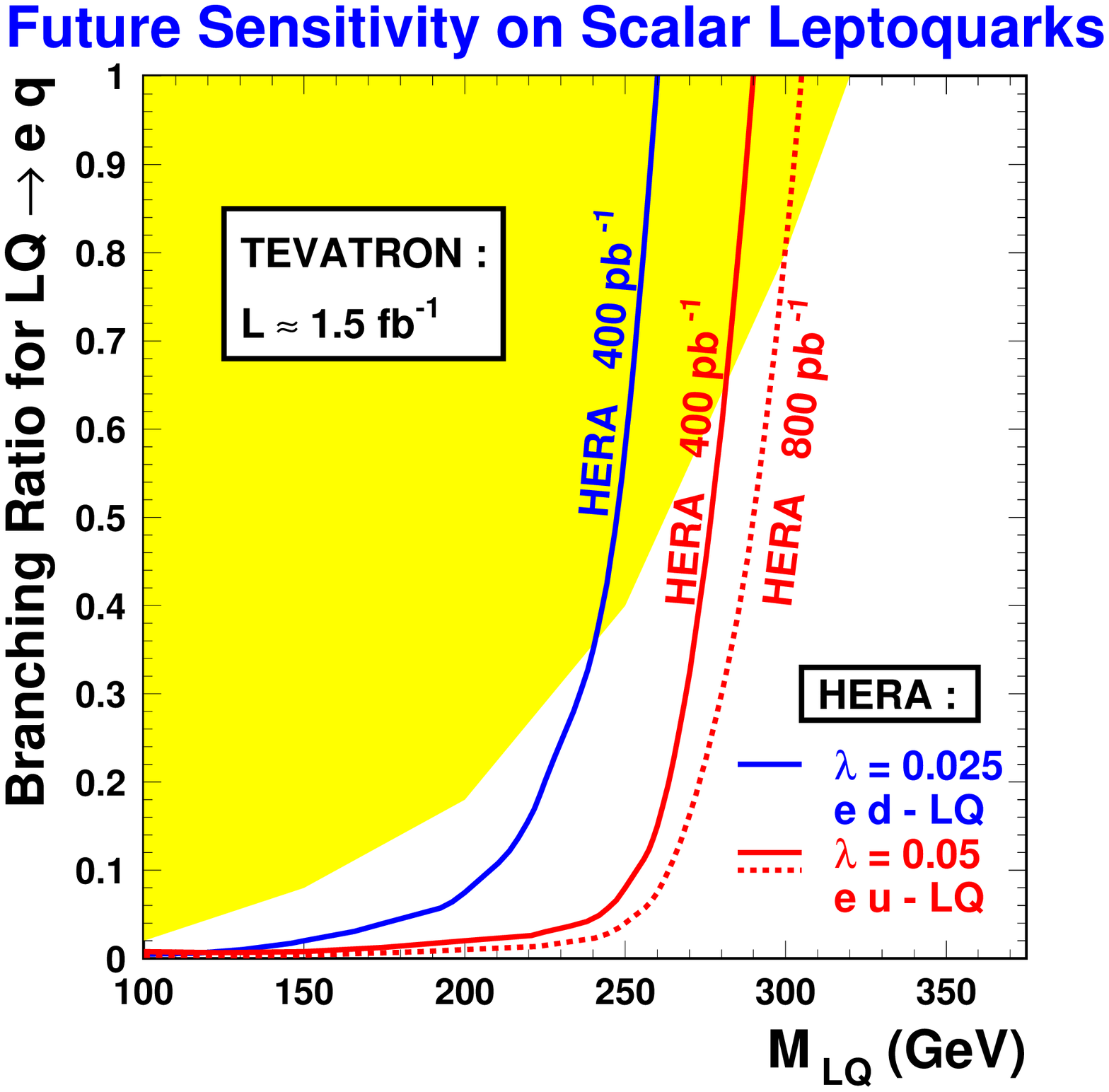,width=9.5cm}
 \caption{\it
      Future prospects for scalar LQ sensitivity for TeVatron Run II
(1.5~fb$^{-1}$) and HERA 2 (three different cases of coupling, LQ species and
luminosity).  Regions to the left of the curves are excluded.
    \label{PEREZ} }
\end{figure}
During the ``HERA 1'' running, both H1 and ZEUS experiments collected
about 110~pb$^{-1}$ of $e^+p$ and 15~pb$^{-1}$ of $e^-p$ data.
Not all of the searches have finished analyzing all of the data yet, but
so far no evidence for new physics has been observed.  Therefore, new
constraints on various physics have been derived: leptoquarks, squarks
in $R$-parity violating SUSY, $eeqq$ contact interactions, large extra
dimensions, quark radius, lepton flavor violation and excited fermions.
Limits are comparable and largely complementary to those obtained in
searches at LEP or the TeVatron.
The excess of isolated leptons in the H1 data is intriguing, although
the ZEUS results from the total data sample are consistent with SM expectations.
Limits on single-top production yield the most stringent constraint on
the FCNC coupling at the $tu\gamma$ vertex.

HERA and both experiments have been performing shutdown work since September 2000
for the luminosity upgrade.  New focusing magnets are installed very
close to the interaction point, even inside the detectors.  The new optics,
together with a moderate increase in beam currents, will bring a
five-fold increase in the instantaneous luminosity compared to the
year 2000 running.  In addition the detectors will be upgraded; for
example, ZEUS will have a micro-vertex detector for the first time.
From summer 2001, the machine will restart and physics running is
scheduled from December.  In the five-year program of ``HERA 2''
running, approximately 1~fb$^{-1}$ of data are expected per experiment.
Another notable feature of the new running is that the longitudinal
polarization of the electron/positron beam will be available by default
to the collider experiments.

In these five years, HERA 2 and TeVatron Run II will be the only
energy-frontier machines in operation.  They will compete in many
physics scenarios.  For example, leptoquark (or squark) sensitivity
is illustrated in Fig.~\ref{PEREZ}\cite{PEREZ}.  If the Yukawa coupling
$\lambda$ is within the sensitivity of HERA, and if HERA 2 quickly starts
up with high-luminosity running, there is a chance for discovery with
10 times more data to come.  Otherwise, for the case of leptoquarks
decaying 100\% to $eq$, with 1-2~fb$^{-1}$ of Run II data the TeVatron
experiments can close the discovery window for direct observation at
HERA since the  $\lambda$-independent mass limit will exceed the
HERA kinematic limit.
However, there are cases where HERA has still potential not covered
by Run II; as indicated in the figure, in the case of leptoquarks with a
small branching ratio to $eq$, HERA's sensitivity exceeds that of Run II
even for fairly small Yukawa coupling.
Also there are some models not probed extensively at the TeVatron
such as excited leptons.
Therefore, the two colliders will continue to be competitive and complementary
during the coming years before the start of LHC,
in the hunt for possible excitement.
\section{Acknowledgements}
The author is very grateful to the organizers of ``XV Rencontres de
Physique de la Vallee d'Aoste'' for their excellent hospitality
which ensured a pleasant stay at La Thuile.
\end{document}